\documentclass{article}

\begin{document}
\title{DIRAC MONOPOLES IN THE ERNST--SCHWARZSCHILD SPACETIME}
\author{ A. A. BYTSENKO$^1$ AND YU. P. GONCHAROV$^2$ \\
\mbox{\small{1. Departamento de Fisica, Universidade Estadual de Londrina}}\\
\mbox{\small{Caixa Postal 6001, Londrina--Parana, Brazil}}\\
\mbox{\small{2. Theoretical Group, Experimental Physics Department, State
          Polytechnical University}} \\ 
\mbox{\small{Sankt--Petersburg 195251, Russia}}}

\maketitle
\begin{abstract}
It is discussed that the Ernst--Schwarzschild metric describing
a nonrotating black hole in the external magnetic field admits the solutions
of the Dirac monopole types for the corresponding Maxwell equations.
The given solutions are obtained in explicit form and a possible influence
of the conforming Dirac monopoles on Hawking radiation is also outlined.

\end{abstract}

\section{Introduction}  

In astrophysics for a long time the physics of black holes immersed into
an external magnetic field has been studied ( see, e.g., the review in Ref. 1).
In view of it the whole class of solutions of the Einstein--Maxwell equations
was found to model a black hole in an external electromagnetic field. Referring
for more details to Refs. 3 we should like here to notice that
an isolated black hole might possess the internal magnetic fields of the Dirac
monopole types. The latter configurations should be connected with nontrivial
topological properties of black holes and could have an essential influence on
quantum processes near black holes, for instance, on Hawking radiation. A 
number
of examples of such configurations may be found in Refs. 3 and 
references therein. Physically, the existence of those configurations should
be obliged to the natural presence of magnetic U(N)-monopoles (with $N\ge1$)
on black holes though the total (internal) magnetic charge (abelian or
nonabelian) of black hole remains equal to zero. One can consider that
monopoles reside in black holes as quantum objects without having influence on
the black hole metrics. They could reside in the form of monopole gas in which
the process of permanent creation and annihilation of the virtual
monopole-antimonopole pairs occurs so that the summed internal magnetic charge
(i. e., related with topological properties) is equal to zero while the 
external
one (not connected with topological properties) may differ from zero 
( e. g. on the Reissner-Nordstr\"om or, more generally, 
Kerr-Newman black holes with magnetic charges).
While existing the virtual monopole-antimonopole pair can interact 
with a particle
and, by this, increasing the Hawking radiation (see Refs. 3 and
references therein).

  There arises the question whether there exist the Dirac-like monopole
configurations on the black holes immersed into an external magnetic field.
Within the given note we shall show that the answer is affirmative by
example of the Ernst--Schwarzschild spacetime~\cite{Er76} describing
Schwarzschild black hole in asymptotically homogeneous magnetic field.
The metric of spacetime manifold in question is
$$
ds^2=g_{\mu\nu}dx^\mu\otimes dx^\nu\equiv
\Lambda^2(adt^2-a^{-1}dr^2-r^2d\vartheta^2)-
\frac{r^2\sin^2\vartheta d\varphi^2}{\Lambda^2} \eqno(1)
$$
with $a=1-2M/r$, $\Lambda=1+\frac{1}{4}B^2r^2\sin^2\vartheta$,
$|g|=|\det(g_{\mu\nu})|=(\Lambda^2r^2\sin\vartheta)^2$
and $0\leq r<\infty$, $0\leq\vartheta<\pi$,
$0\leq\varphi<2\pi$. At this the surface $t$=const, $r$=const is an ellipsoid
with topology ${\cal{S}}^2$.

  Throughout the paper we employ the system of units with $\hbar=c=G=1$,
unless explicitly stated.

\section{Dirac Monopole Type Solutions}

To write down the Maxwell equations in spacetime with metric (1) we need to
know the action of the Hodge star operator * on 2-forms
$F=F_{\mu\nu}dx^\mu\wedge dx^{\nu}$ which is defined for any $k$-dimensional
(pseudo)riemannian manifold
$B$ provided with a (pseudo)riemannian metric $g_{\mu\nu}$ by the relation
(see, e. g., Refs. 5)
$$
F\wedge\ast F=(g^{\mu\alpha}g^{\nu\beta}-g^{\mu\beta}g^{\nu\alpha})
F^a_{\mu\nu}F^a_{\alpha\beta}
\sqrt{|g|}\,dx^1\wedge dx^2\cdots\wedge dx^k 
\eqno(2)
$$
in local coordinates $x^\mu$. In the case of the metric (1) this yields
for the basis elements
\begin{eqnarray*}
\ast(dt\wedge dr)& = & \sqrt{|g|}g^{tt}g^{rr}d\vartheta\wedge d\varphi=
-\frac{r^2\sin\vartheta}{\Lambda^2}d\vartheta\wedge d\varphi\>,  \\ 
\ast(dt\wedge d\vartheta) & = & -\sqrt{|g|}g^{tt}g^{\vartheta\vartheta}
dr\wedge d\varphi=
\frac{\sin\vartheta}{a\Lambda^2}dr\wedge d\varphi\>,   \\
\ast(dt\wedge d\varphi) & = & \sqrt{|g|}g^{tt}g^{\varphi\varphi}
dr\wedge d\vartheta=-\frac{\Lambda^2}{a\sin\vartheta}dr\wedge d\vartheta\>,  \\
\ast(dr\wedge d\vartheta) & = & \sqrt{|g|}g^{rr}g^{\vartheta\vartheta}
dt\wedge d\varphi=\frac{a\sin\vartheta}{\Lambda^2}dt\wedge d\varphi\>,  \\
\ast(dr\wedge d\varphi) & = & -\sqrt{|g|}g^{rr}g^{\varphi\varphi}
dt\wedge d\vartheta=-\frac{a\Lambda^2}{\sin\vartheta}dt\wedge d\vartheta\>,  
\end{eqnarray*}
$$
\ast(d\vartheta\wedge d\varphi) =  \sqrt{|g|}
g^{\vartheta\vartheta}g^{\varphi\varphi}dt\wedge dr=
\frac{\Lambda^2}{r^2\sin\vartheta}dt\wedge dr\>, 
\eqno{(3)}
$$
so that $\ast^2=\ast\ast=-1$, as should be for the manifolds with lorentzian
signature~\cite{Bes87}.

The Maxwell equations are
$$
dF=0, 
\eqno{(4)}
$$
$$
d*F=0 
\eqno{(5)}
$$
for electromagnetic vector-potential $A=A_{\mu}dx^\mu$, $F=dA$
with the exterior differential $d=\partial_t dt+\partial_r dr+
\partial_\vartheta d\vartheta+\partial_\varphi d\varphi$ in coordinates
$t,r,\vartheta,\varphi$.
It is clear that (4) is identically satisfied (the Bianchi identity)
so that it is necessary to solve only the Eq. (5).
Let us search for $A$ in the form $A=A_\varphi(r,\vartheta) d\varphi$,
i.e., putting the components $A_t=A_r=A_\vartheta=0$. This entails
$F=dA=\partial_rA_\varphi dr\wedge d\varphi
+\partial_\vartheta A_\varphi d\vartheta\wedge d\varphi$
and with the help of (3)
$$
\ast F=
-\frac{a\Lambda^2}{\sin\vartheta}\partial_rA_\varphi dt\wedge d\vartheta
+\frac{\Lambda^2}{r^2\sin\vartheta}\partial_\vartheta A_\varphi dt\wedge dr.
\eqno(6)
$$
Then the Eq. (5) take the form
$$
\frac{\partial}{\partial r}\left[\sqrt{|g|}\left(
g^{rr}g^{\varphi\varphi}\frac{\partial A_\varphi}{\partial r}\right)\right]
+\frac{\partial}{\partial \vartheta}\left[\sqrt{|g|}\left(
g^{\vartheta\vartheta}g^{\varphi\varphi}
\frac{\partial A_\varphi}{\partial \vartheta}\right)\right]=
$$
$$\frac{\partial}{\partial r}\left[\frac{a\Lambda^2}{\sin\vartheta}
\frac{\partial A_\varphi}{\partial r}\right]
+\frac{\partial}{\partial \vartheta}\left[\frac{\Lambda^2}{r^2\sin\vartheta}
\frac{\partial A_\varphi}{\partial \vartheta}\right]=0 \, .
\eqno{(7)}
$$
Now we employ the ansatz $A_\varphi=-\alpha f(\vartheta)/\Lambda$ with some
constant $\alpha$ and
inserting it into (7) entails the equation for function $f(\vartheta)$
$$
\sin\vartheta\frac{d^2f}{d^2\vartheta}
-\cos\vartheta\frac{df}{d\vartheta}=0 \, .
\eqno{(8)}
$$
The solution of (8) necessary to us is $f(\vartheta)=\cos\vartheta$ so
$$
A= -\frac{\alpha \cos\vartheta}{\Lambda}d\varphi \, .
\eqno{(9)}
$$
To fix the constant $\alpha$ let us require the fulfillment the Dirac charge
quantization condition
$$
\int\limits_{{\cal{ S}}^2}\,F=
\int\limits_{{\cal{ S}}^2}\,\partial_\vartheta A_\varphi d\vartheta\wedge d\varphi=
4\pi q= 4\pi\frac{n}{e}
\eqno{(10)}
$$
with magnetic charge $q=n/e$, $n\in {\cal{ Z}}$, the set of integers, where we
integrate over any
surface $t$=const, $r$=const with topology ${\cal{ S}}^2$, $e$ is elementary
electric charge.
The direct evaluation gives
$$
\int\limits_{{\cal {S}}^2}\,F=2\pi\alpha\int_0^\pi\frac{\sin\vartheta}{\Lambda}
\left(1+\frac{B^2r^2\cos^2\vartheta}{2\Lambda}\right)d\vartheta=4\pi\alpha,
\> \eqno(11)
$$
so that $\alpha=n/e$. Also it is not complicated to check that the Gauss
theorem holds true
$$
\int_{{\cal{S}}^2}\ast F=0.\> \eqno(12)
$$
One can notice that relation (9) passes on to the conforming one for
the case of the pure Schwarzschild metric, i.e. at $B=0$~\cite{GF}. Finally 
it is easy to check that the given solutions satisfy the Lorentz gauge
condition that can be
written in the form ${\rm div}(A)=0$, where the divergence of 
1-form $A=A_\mu dx^\mu$ is defined by the relation
$
{\rm div}(A)=\frac{1}{\sqrt{|\delta|}}\partial_\mu(\sqrt{|\delta|}g^{\mu\nu}
A_\nu) \, .$

\section{Concluding Remarks}

Mathematical reason for existence of the solutions obtained is the following
one. It should be noted that the standard spacetime topology on which
the metric (1) with arbitrary $a=a(r)$ can be realized in a natural way is of
bh-form. As was discussed in Refs. 3, such topology admits
countable number of complex line bundles
while each complex line bundle $E$ can be characterized by its
Chern number $n\in{\cal{Z}}$. The solutions obtained are just connections in
the mentioned bundles -- Dirac monopoles.
But it should be emphasized that the total
(internal) magnetic charge $Q_m$ of system(black hole + external magnetic
field) which
should be considered as the one summed up over all the monopoles remains
equal to zero because
$$
Q_m=\frac{1}{e}\sum\limits_{n\in{\cal{Z}}}\,n=0 \, , 
\eqno{(13)}
$$
so the external observer does not see any magnetic charge of system
though the monopoles are present in the sense described above.

On the other hand, the nontrivial topological properties of spacetimes
may play essential role while studying quantum geometry of fields on them
(see, e. g., our reviews~\cite{GB}). Therefore physically the results obtained
could mean that in the given spacetime
there exist topologically inequivalent configurations (TICs) for various fields.
Each TIC corresponds to its Chern number
$n\in \cal{Z}$. TIC with $n=0$ can be called {\it untwisted},
while the rest of the TICs with $n\not=0$ should be referred to as
{\it twisted}. For example, TICs of complex scalar field $\phi$ with mass
$\mu_0$ should obey the equations
$$
|g|^{-1/2}
(\partial_\mu-ieA_\mu)[g^{\mu\nu}(|g|)^{1/2}
(\partial_\nu-ieA_\nu)\phi]=-\mu_0^2\phi \, ,
\eqno{(14)}
$$
where along with external electromagnetic field
$A=-B^2r^2\sin^2\vartheta/(2\Lambda)d\varphi$ we should include the addendum
corresponding to (9) so that the full $A$ of (14) would be of the form
$$
A=-\left(\frac{B^2r^2\sin^2\vartheta}{2\Lambda}
+\frac{n\cos\vartheta}{e\Lambda}\right)d\varphi \, . 
\eqno{(15)}
$$
Analogously this also holds true for spinor field.

  Under the circumstances one can speak about the Hawking radiation
process for any TIC of complex scalar or spinor fields and one may try
to get the luminosity $L(n)$ with respect to the Hawking radiation for TIC
with the Chern number $n$.
We can interpret $L(n)$ with $n\ne0$ as an additional contribution to the
Hawking radiation due to the additional charged particles leaving
the black hole because of the interaction with monopoles and
the conforming radiation
can be called {\it the monopole Hawking radiation}~\cite{Gon99}. 
Under this situation,
for the all configurations luminosity $L$ of black hole in question with
respect to the Hawking radiation to be obtained,
one should sum up over all $n$, i. e.
$$
L=\sum\limits_{n\in{\cal{Z}}}\,L(n) \, . \eqno{(16)}
$$
As a result, we can expect a marked increase of Hawking radiation from
black holes under consideration.
 The above program has been to a large extent realized for Schwarzschild black
holes~\cite{{GF},{GF01}} and there is an interest to look at how the results
obtained before would be changed in the presence of external magnetic field.

But for to get an exact value of this increase one should
apply numerical methods. In the case of the pure Schwarzschild black hole,
for example, it was found that the contribution due to monopoles
can be of order 11~\%  of the total pion-kaon luminosity~\cite{GF} while
it is of order 22 \% for electron-positron case~\cite{GF01}. So that
it would be interesting enough to evaluate similar increase in the case
of the Ernst--Schwarzschild metric.

\medskip
\noindent{\bf Acknowledgements} 
\medskip

    The work of authors was supported in part by the Russian
Foundation for Basic Research (grant No. 01-02-17157).

\end{document}